\documentclass[conference]{IEEEtran}
\IEEEoverridecommandlockouts
\usepackage[utf8]{inputenc}
\usepackage{cite}
\usepackage{amsmath,amssymb,amsfonts}
\usepackage{algorithmic}
\usepackage{graphicx}
\usepackage{tikz}
\usepackage{textcomp}
\usepackage{xcolor}
\usepackage[hyphens]{url}
\usepackage[hidelinks]{hyperref}
\usepackage[all]{hypcap}
\definecolor{orange}{HTML}{FC8D59}
\definecolor{yellow}{HTML}{FFFFBF}
\definecolor{newblue}{HTML}{91BFDB}

\usetikzlibrary{arrows.meta}
\tikzset{%
	>={Latex[width=2mm,length=2mm]},
	base/.style = {rectangle, rounded corners, draw=black,
		minimum width=1cm, minimum height=0.4cm,
		text centered, font=\sffamily},
	user/.style = {base, fill=gray!20},%user/.style = {base, fill=gray!20},
	publicInformation/.style = {base, fill=yellow!60},%publicInformation/.style = {base, fill=red!30},
	privateInformation/.style = {base, fill=orange!60},%privateInformation/.style = {base, fill=yellow!30},
	circle/.style = {base, fill=newblue!60},%circle/.style = {base, fill=newblue!30},
	empty/.style = {draw=none},
	pics/vhsplit/.style n args = {5}{
		code = {
			\node[text width=0.5cm] (A) at (#1) {};
			\node[anchor=south west,text width=3cm] (B) at (A.east) {#3};
			\node[anchor=north west,text width=3cm] (C) at (A.east) {#4};
			\node[inner sep=0pt,draw,rounded corners,fit=(A)(B)(C),blend mode=overlay,overlay,fill=#5] (outer) {}; 
			\node[xshift=-1.6cm] (tt) at (outer.center) {#2};  
			\draw (B.north west) -- (C.south west)
			(B.south west) -- (C.north east);    
		}
	},
	pics/vhsplit2/.style n args = {5}{
		code = {
			\node[text width=0.5cm] (A) at (#1) {};
			\node[anchor=south west,text width=3.5cm] (B) at (A.east) {#3};
			\node[anchor=north west,text width=3.5cm] (C) at (A.east) {#4};
			\node[inner sep=0pt,draw,rounded corners,fit=(A)(B)(C),blend mode=overlay,overlay,fill=#5] (outer) {}; 
			\node[xshift=-1.85cm] (tt) at (outer.center) {#2};  
			\draw (B.north west) -- (C.south west)
			(B.south west) -- (C.north east);    
		}
	}
}

\usepackage{epsfig}
\usepackage{calc}
\usepackage{amstext}
\usepackage{fontawesome}
\def\BibTeX{{\rm B\kern-.05em{\sc i\kern-.025em b}\kern-.08em
    T\kern-.1667em\lower.7ex\hbox{E}\kern-.125emX}}
\begin{document}

\title{A Proxy-Based Encrypted Online Social Network With Fine-Grained Access\thanks{The authors acknowledge the financial support by the Federal Ministry of Education and Research of Germany in the framework of SoNaTe (project number 16SV7405).}}

\author{\IEEEauthorblockN{Fabian Schillinger}
\IEEEauthorblockA{\textit{Computer Networks and Telematics}\\\textit{Department of Computer Science}\\\textit{University of Freiburg}\\
Freiburg, Germany\\
\href{https://orcid.org/0000-0001-8771-8290}{\includegraphics[height=1.3\fontcharht\font`\B]{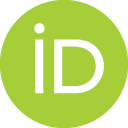}\, \textcolor{blue}{\url{https://orcid.org/0000-0001-8771-8290}}}
%\href{}{\includegraphics[height=1.3\fontcharht\font`\B]{ORCID-iD_icon-128x128}\, }
}\and
\IEEEauthorblockN{Christian Schindelhauer}
\IEEEauthorblockA{\textit{Computer Networks and Telematics}\\\textit{Department of Computer Science}\\\textit{University of Freiburg}\\
Freiburg, Germany\\
\href{https://orcid.org/0000-0002-8320-8581}{\includegraphics[height=1.3\fontcharht\font`\B]{ORCID-iD_icon-128x128}\, \textcolor{blue}{\url{https://orcid.org/0000-0002-8320-8581}}}}
}

\maketitle

\begin{abstract}
When using Online Social Networks, users often share information with different social groups. When considering the backgrounds of the groups there is often no or little intersection within the members. This means that a user who shares information often has to share it with all members of all groups. It can be problematic that the user cannot decide which group sees which information. Our approach therefore, allows users to decide for every bit of information who can access it. Further, protected circles can be created, where users can share information within. Shared information and circles are encrypted and the keys can be distributed by proxies.
\end{abstract}


\section{Introduction}
Online Social Networks (OSNs) experience strong popularity and the number of people with at least one social media account is increasing steadily. Users of OSNs provide information about themselves and about things they like and dislike. As this information is arranged by the user itself it paints a specific image about the user. This is certainly in the interest of the user if he is among his own group, but it may be disadvantageous or even inappropriate if the user is part of different groups. E.g. if the user shares information with friends about some shared hobby, while colleagues of the user also participate on the same platform and can see this information. Therefore, being able to define fine-grained policies for all shared information on who can access it and protecting this information from being accessed by other users is an important target, which we address with our scheme.
\subsection*{Our Contribution}
Our contribution is an Online Social Network where information can be made publicly available or protected by encryption. For each protected information the access group can be defined from scratch, such that different users can have the right to access them. Further, social groups can form circles where information is, again, encrypted and thus protected. Keys can be shared with groups of proxies. They can, on the one hand, forward the decryption keys, according to the definitions of the user. On the other hand, they can protect keys together: comparable to secret sharing schemes either all proxies have to provide their part or a predefined portion of them, which has to be approximately met. Additionally, the encryption parameters are different for all users and can be changed, which allows voiding previously granted access.
\subsection*{Organization of the Paper}
The paper is structured as follows: In Section~\ref{Related Work} related work, regarding encrypted and privacy preserving Online Social Networks are discussed. In Section~\ref{Proposed System} the fundamental cryptographic methods for our work are explained and our proposed system is described in detail. Finally, Section~\ref{Conclusion} concludes the work and gives an outlook on possible extensions.
\section{Related Work}
\label{Related Work}
\textit{Safebook}~\cite{5350374} is a decentralized OSN. Social networks, that are present in real life are used to construct trust relationships. The internet is used as the communication and transport layer. A peer-to-peer layer is used to implement application services, such as lookup. The third layer is the social network layer, which is centered on the user. Users are arranged and connected, such that they form concentric structures to provide data storage and communication privacy. These structures protect the centered nodes because connections from outer rings follow trust relations from real life. The nodes on the rings provide profile retrieval and communication obfuscation.\\
\textit{Persona}~\cite{baden2009persona} is another privacy-preserving OSN. Users define who can access their information. Attribute-based encryption is used, such that users can define fine-grained access policies. Further, attribute-based encryption allows sharing ciphertext within groups of participants that share at least one attribute. Additionally, each user has a key-pair for encryption of symmetric keys. The public key is distributed to other users to encrypt data in groups of multiple users. This is an additional method to the attribute-based encryption.\\
Another OSN, utilizing end-to-end encryption is \textit{Snake}\cite{barenghi2014snake}. The OSN is a client-side application, written in HTML5 and JavaScript. The encryption procedures of Snake use the WebCrypto API and are performed only on the client-side. This allows the server to be used as a CRUD interface for ciphertexts, without additional functionality. The design of the database does not allow concluding which users communicate, because addresses are masked.\\
\textit{LotusNet}~\cite{aiello2012lotusnet} is a framework that allows using the peer-to-peer paradigm with strong user authentication in OSNs. A flexible and fine-grained access control system balances security, privacy, and services in OSNs. Distributed hash tables are used as the underlying storage. Overlay nodes provide strong authentication and enable communication with a service layer. Content is stored as encrypted ciphertexts in the DHT layer. Ciphertexts are tagged with arbitrary labels from the users. The search engine finds appropriate data using these labels.\\
In~\cite{zhou2008preserving} neighborhood attacks are analyzed. It is stated, that attackers with knowledge about neighbors of a victim can re-identify them in the OSN. In the first step of the proposed solution, the neighborhoods of all vertices are extracted. In the second step, they are grouped. Neighborhoods in the same groups are anonymized by generalizing vertex labels and inserting additional edges.\\
In~\cite{tai2011privacy} a comparable attack, called friendship attack is described. Here, an adversary uses the degree of two vertices connected by an edge to re-identify victims. $k^2$-degree anonymity is introduced to limit the probability of re-identifying victims. $k^2$-degree anonymity means, that for every edge where the connected nodes have degree $d_1$ and $d_2$ at least $k-1$ other vertices with the same degrees $d_1, d_2$ exist. This is achieved by introducing additional edges into the graph.\\
An access control model for OSNs is presented in~\cite{fong2009privacy}. The model formalizes and generalizes the privacy preservation mechanism of Facebook. The model can be instantiated to express policies which are currently not supported by Facebook and gives a formal framework for the analysis of such policies in other OSNs.\\
In~\cite{cutillo2009privacy} decentralization is used to construct a privacy-preserving OSN. Centralization is seen as the key privacy issue because a potentially malicious service provider is in control of all data. The OSN leverages trust relationships of social networking. A proposed anonymization technique uses multi-hop routing between trusted nodes to provide privacy and data access and exchange.\\
A paradigm to allow users to choose so-called suites of privacy settings is presented in~\cite{bonneau2009privacy}. The suites can be specified by friends of the users or trusted experts and modified. They allow implementing expressive and usable privacy controls. This is considered a major challenge, as a wrong understanding of privacy settings could lead to unwanted disclosure of information in an OSN.\\
In~\cite{feltprivacy} a privacy-by-proxy design for APIs in OSNs is proposed. Often, third-party content is integrated into sites in OSNs. Therefore, third-party developers may get access to user data, which introduces privacy risks. In the work, popular Facebook applications are analyzed and a privacy-preserving API is proposed. It provides anonymized social graphs and placeholders for user data.\\
\textit{FlyByNight}~\cite{Lucas:2008:FMP:1456403.1456405} is an extension to Facebook. It uses client-side JavaScript to encrypt content, such that plaintext data is not sent to Facebook. The keys are managed through the existing infrastructure. An additional password is used to encrypt a private key. This key is stored as a ciphertext on a flyByNight server. The application allows encrypting data for multiple participants. This is achieved by using the public keys of participants to encrypt the data. Public information is encrypted using proxy-cryptography: unique keys are used to encrypt the data. The ciphertexts are modified for every user by the server, such that decryption is possible with their private keys.\\
\textit{None of Your Business (NOYB)}~\cite{Guha:2008:NPO:1397735.1397747} is another extension for Facebook. Here, information is masked. To mask data, like public posts or personal information, it is split into small parts, called atoms. Then, atoms are replaced by other atoms, according to different public dictionaries of atoms from participants and non-participants. The dictionary lookup procedure uses the index of an atom, which then is encrypted with a symmetric key and a random nonce. This ciphertext is used as a new index to find a masking atom. Different dictionaries are used, such that meaningful atoms are used for masking. E.g. the location of a user is masked with another users location.\\
%?\cite{robison2012private}\\
%\cite{kikuchi2004secure}\\
%\cite{yang2008design}\\
%\cite{1458584}\\
%\cite{OMEMO1}\\
%\cite{OTR1}\\
%\cite{SIGNAL_doubleratchet_1}\\
%\cite{rfc4880}\\
%\cite{rfc5321}\\
%\cite{rfc5751}\\
%\cite{rfc6120}\\
%\cite{telegram_2018}\\
%\cite{moscaritolo2012silent}\\
\section{Proposed Online Social Network}
\label{Proposed System}
For the encryption of small content the asymmetric scheme of Elgamal~\cite{1057074} is used. A message $m$ is encrypted for a recipient $B$ by the sender $A$ as follows: The sender $A$ chooses a number $r$ uniformly random between 0 and $p-1$, where $p$ is a large prime and $p-1$ has at least one large prime factor. $A$ computes ${c_0 \equiv g^r\mod p}$ and ${c_1\equiv m\cdot g^{x_Br}\mod p}$, where $g$ is a public value, known to both $A$ and $B$. $g^{x_B}\mod p$ is the public key of $B$. The tuple $(c_0, c_1)$ is the ciphertext. $B$ calculates ${g^{x_Br}\equiv (c_0)^{x_B}\mod p}$, where $x_B$ is the private key of $B$. By computing ${c_1/g^{x_Br}\equiv m\cdot g^{x_Br}/g^{x_Br}\equiv m\mod p}$ the message is decrypted.\\ The scheme allows to encrypt a message for multiple recipients $B, C, \dots$ with public keys $x_B, x_C, \dots$. The ciphertext ${(c'_0, c'_1) \equiv (g^r,m\cdot  g^{r(x_B+x_C+\dots)}\mod p}$ can be decrypted, when all recipients compute ${d \equiv (c'_0)^{(x_B+x_C+\dots)} \mod p}$ together. The message is computed, again by dividing $c'_1$ by $d$: $m \equiv c'_1/d\mod p$. The random parameter of a ciphertext ${(c_0, c_1) \equiv (g^r, g^{Xr}\cdot m) \mod p}$ can be modified by anybody with knowledge of the public key $g^X$. No knowledge of the current random parameter $r$, message $m$, or private key $x$ is needed. By computing $c'_0 \equiv c_0\cdot g^q\mod p$ and $c'_1 \equiv c_1 \cdot (g^X)^q\mod p$ the ciphertext is modified to ${(c'_0, c'_1) \equiv (g^{r+q}, m\cdot g^{X(r+q)}) \mod p}$. The encrypted message $m$ stays the same. These two properties are utilized by our approach.
\subsection{Overview}
\begin{figure*}[t]
	\centering
	\begin{tikzpicture}[every node/.style={font=\sffamily}, align=center]

	\node (user) [user] at (0,0) {\faUser~User};
	
	\node (personalInformation) [publicInformation] at (4.4,2.0) {Public Information};
	\node (publicinformation1) [publicInformation] at (5.4,0.9) {Name};
	\node (publicinformation2) [publicInformation] at (3.2,0.9) {Residence};
	\node (publicinformation3) [empty] at (7.0,0.9) {\dots};
	\node (weather) [publicInformation] at (4.4,0.3) {Post about the weather};	
	
	\node (circles) [circle] at (4.4,-0.7) {Circles};
	\node (circlesLock1) [empty] at (3.4,-1.8) {\faKey\textsubscript{c\textsubscript{0}}};
	\node (circlesLock2) [empty] at (4.4,-1.8) {\faKey\textsubscript{c\textsubscript{1}}};
	\node (circlesLock3) [empty] at (5.4,-1.8) {\dots};
	
	\node (circlesEmpty1) [circle] at (2.4,-2.9) {Content};
	\node (circlesEmpty2) [empty] at (4.1,-2.9) {\dots};

	\node (privateInformation) [privateInformation] at (-5,1) {Private Information};
	\node (privateInformationLock1) [empty] at (-5.8,-0.1) {\faKey\textsubscript{pi\textsubscript{0}}};
	\node (privateInformationLock2) [empty] at (-5,-0.1) {\faKey\textsubscript{pi\textsubscript{1}}};
	\node (privateInformationLock3) [empty] at (-4.2,-0.1) {\dots};
	
	\node (political) [privateInformation] at (-7.2,-1.7) {Post about\\ the political\\ situation};
	
	\node (dob) [privateInformation] at (-4.8,-2.2) {Date of Birth};
	\node (pn) [privateInformation] at (-1.6,-2.2) {Phone Number};
	\node (pidots) [empty] at (-3.1,-2.8) {\dots};
	
	\draw[->] (user) -- (personalInformation.south west);
	\draw[->] (user) -- (circles.north west);
	\draw[->] (user) -- (privateInformation.south east);
	
	\draw[->] (privateInformation) -- (privateInformationLock1.north);
	\draw[->] (privateInformation) -- (privateInformationLock2.north);
	\draw[->] (privateInformation) -- (privateInformationLock3.north);
	
	\draw[->] (circles) -- (circlesLock1.north);
	\draw[->] (circles) -- (circlesLock2.north);
	\draw[->] (circles) -- (circlesLock3.north);
	
	\draw[->] (personalInformation) -- (publicinformation1);
	\draw[->] (personalInformation) -- (publicinformation2);
	\draw[->] (personalInformation) -- (publicinformation3);
	
	\draw[->] (personalInformation) -- (weather.north);
	
	\draw[->] (privateInformationLock2) -- (dob.north);		
	\draw[->] (privateInformationLock2) -- (pn.north);			
	\draw[->] (privateInformationLock2) -- (pidots.north);
	
	\draw[->] (privateInformationLock1) -- (political.north);
	
	\draw[->] (circlesLock1) -- (circlesEmpty1.north);
	\draw[->] (circlesLock1) -- (circlesEmpty2.north);
	
	\end{tikzpicture}
	
	\caption{A user can share private and public information. For every information the user can define who is allowed to access it. Further, circles can be joined. In circles, users can share information with all members of the same circle. Circles and private information are secured by encryption.}
	\label{fig:model}
\end{figure*}
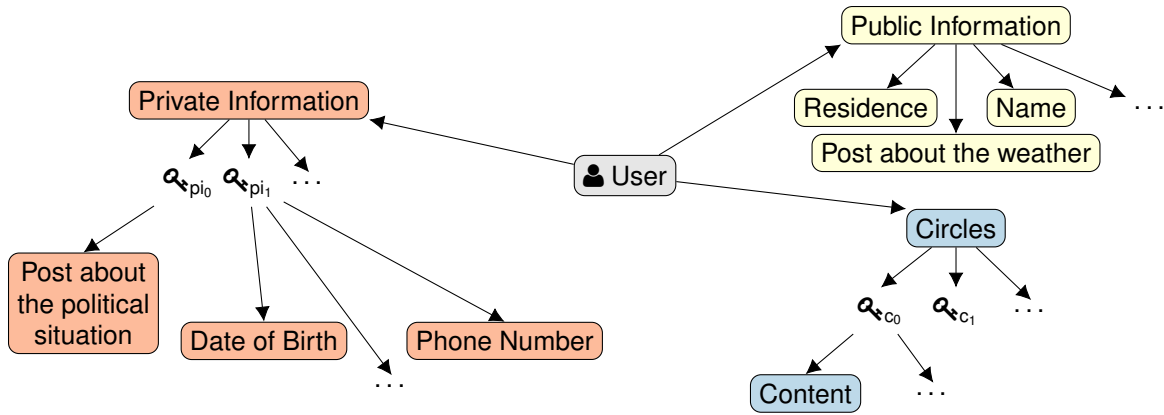

In the proposed OSN each user can share public and private information with other users. This, for example can be the name of the user, the employment status, the phone number, or posts. A user can decide which information is public and which information is private. E.g. it is possible, that a post about the current political situation is marked private by the user, whereas another post about the weather may be marked as public. Private information is encrypted, and therefore can only be accessed by specific users. The user can decide which group of users can decrypt the information by sharing the key with them, or a group of proxies. The proxies can relay the key when the user is not available. Users can join circles. Circles are groups of users that share information with all other members of the circle. Circles and private information are encrypted to ensure privacy. Most likely, private information is small in size, then it is encrypted using the asymmetric scheme of Elgamal. For different parts of information different keys are generated. When private information is too large, it is encrypted using a faster symmetric scheme. Then, the corresponding key is encrypted using the scheme of Elgamal. Information in circles, always is encrypted using a common symmetric key. This key is shared, again as a ciphertext, created with the scheme of Elgamal. This key is can be distributed by proxies. Each proxy receives at least one share of the key. A user, that is new to the circle has to receive every share of the key from the proxies. The model is depicted in Figure~\ref{fig:model}.
\subsection{Encryption and Decryption of Content}
Content is distinguished into two types: public content and protected content. The user can decide, which of his content is public or protected. Public content can be accessed by all users of the OSN. Protected content is either private information or circle content. Private information is content, which is shared by a user with specific groups of users, e.g. specific friends. Parts of private information, often, are short and therefore are encrypted, using the Elgamal encryption scheme. No noteworthy differences in the run-times, when compared to symmetric encryption algorithms should be noticeable. Larger content, such as circle content, on the other hand, is encrypted, using a symmetric encryption algorithm. The necessary keys are short and therefore encrypted using the Elgamal encryption scheme. All protected content, which is encrypted, using the Elgamal scheme can be shared through a proxy system. This content in the following is denoted as short content.\\
Short content $m$ can be encrypted using multiple public proxy-keys $g^a, g^b, \dots$ of users $a, b,\dots$ and a randomly chosen encryption parameter $r$. By computing \begin{equation}\label{eq:encrypt content}
\begin{split}
c & \equiv m\cdot {(g^a)}^r\cdot {(g^b)}^r \cdot \dots \\
& \equiv m\cdot g^{r(a+b+\dots)}\mod p
\end{split}
\end{equation} content is encrypted. $c$ and $g^r$ are sent to the server. This allows the server to distributed short content. When a user wants to access content $c$, the server re-encrypts it. For the re-encryption procedure the server generates two random numbers $s$ and $t$. Then, $c$ is modified for $u$ by calculating 
\begin{equation}
\begin{split}
c_u & \equiv c \cdot t \\
& \equiv m \cdot g^{r(a+b+\dots)}\cdot t \mod p .
\end{split}
\end{equation} 
Further, using the public keys $g^a, g^b, \dots$ of the according users $a,b,\dots$ the value $p_u$ is calculated as 
\begin{equation} 
\begin{split}
p_u & \equiv t \cdot {(g^a)}^s \cdot {(g^b)}^s \cdot \dots \\
& \equiv t\cdot g^{s(a+b+\dots)} \mod p. 
\end{split}
\end{equation} 
The server sends $c_u, p_u, g^s$, and $g^r$ to $u$. To decrypt the short content $u$ has to send $g^r$ and $g^s$ to all users $a, b, \dots$. They then can compute the decryption values ${(g^r)}^{-a}, {(g^r)}^{-b}, \dots$ and ${(g^s)}^{-a}, {(g^s)}^{-b}, \dots$. Then, $u$ can compute $m\cdot t$ as: 
\begin{equation}
\begin{split}
m\cdot t & \equiv c_u\cdot g^{-ar} \cdot g^{-br}\cdot \dots \\
& \equiv m\cdot t\cdot g^{r(a+b+\dots)} \cdot g^{-r(a+b+\dots)} \mod p
\end{split}
\end{equation} 
and $t$ as: 
\begin{equation}
\begin{split}
t & \equiv p_u \cdot g^{-as} \cdot g^{-bs} \cdot \dots \\
& \equiv t\cdot g^{s(a+b+\dots)}\cdot g^{-s(a+b+\dots)}  \mod p.
\end{split}
\end{equation} 
To decrypt $m$ the user has to compute $\frac{m\cdot t}{t}\mod p$. It is possible to encrypt short content with a single proxy-key. The calculations stay similar, but only one shareholder $s$ has to calculate the decryption values for $c_u$ and $p_u$. \\The benefit of the additional factors $s$ and $t$ is, that the server can delete them, when the proxies request it. When $u$ wants to receive $c_u$ again from the server it is a new value and cannot be decrypted with the values $u$ received the previous time. This allows revoking the access to resources.
\subsection{Distribution of Proxy-Keys}
Proxy-keys are used to distribute short content. Short content can be decrypted, only if all contributors participate. In the simple case, a proxy-key $p^a$ is just the public key of user $A$. As public keys are stored on the server, everybody can use them for the distribution of data. Only user $A$ knows the corresponding private key $a$ and can compute $g^{-a}$, which is needed for the decryption.\\ In the other case, a proxy-key $g^x$ is generated by choosing $x$ at random. $g^x$, then, is encrypted for users $A, B, \dots$. This is achieved by choosing random parameters $r_a, r_b, \dots$ and calculating tuples $(g^{r_a}, g^{ar_a} \cdot x), (g^{r_b}, g^{br_b} \cdot x), \dots$, where $g^a, g^b,\dots$ are the public keys of users $A,B,\dots$. This allows distributing the same proxy-key to multiple users. This is used for the distribution of private information and circle content.\\ A user specifies for each private information the group of users, which is allowed to decrypt the information. For example, the user $u$ may share some information $i$, like his date of birth and phone number, with all friends $f \in F$, whereas $u$ wants to share other information $j$, like his residence, with all friends and their friends $G$. The user, then, can encrypt $i$ using a key $x$. Content $j$ is encrypted using the key $y$. The key $y$ can be shared with all $f \in F$. If another user $g \in G$ wants to decrypt such content, every $f$ can help him decrypt it. The other key $x$ can be shared with all users $f$, such they can decrypt it. But, there is no need for $u$ to share $x$ or $y$ with all $f \in F$, but with some $f' \in F' \subset F$, such that users $f^* \in F\setminus F'$ can receive them from any user $f'$.\\ The keys for circle content can be shared with a quorum $Q$ of participants in the circle. Then, for example, three proxy-keys $a$, $b$, and $c$ are generated and used to encrypt the content, as described in Equation~\ref{eq:encrypt content}. The keys, then, are distributed to $Q$, such that each member ${q \in Q}$ receives one key. This allows users to decrypt the content if three members with mutually different keys help in the decryption procedure.\\Members of the quorum can receive multiple keys, such that different probabilistic thresholds can be achieved. One way of distributing these keys to the members is to hand a key to every member, such that they are equally distributed. Every following key for the members then is randomly chosen from all keys which are not already distributed to the member. When applying this scheme to a quorum of size 10, with 6 different keys ${k \in (0,1,\dots,5)}$ and two keys per member ${m\in (0,1,\dots,9)}$, a member $m$ receives the keys ${i = m\mod 6}$ and ${j \neq i}$. Simulations show, that a probabilistic threshold of 5 can be achieved using this distribution: on average one has to pick 5.5 members to receive all keys and the probability that 4, 5, or 6 picks are needed is about $0.75$.
\section{Conclusion}
\label{Conclusion}
We introduced a scheme for a partially encrypted Online Social Network. Users can define, which information is protected through encryption, and who can decrypt it. Information can be shared with different groups of users. Further, circles of users can be created where the content is encrypted, too. The keys can be distributed by a proxy system. This allows the sharing of keys, even if the user is not available. Another benefit of proxies is, that each proxy can receive one or more different portions of a key. This allows granting access in different scenarios: all proxies are needed or an approximate threshold of collaborating proxies is defined. The keys are modified by the server for each user. The parameters of this modification can be deleted, which allows the voiding of access. \\Further improvements to the scheme can be made by applying a real secret sharing scheme for the portions of the keys. This allows for creating definite threshold values. Additional modifications could move the modification of the keys from the server to the proxies. This reduces the ability of the server to mischievously erase encryption parameters for users.

\bibliography{quorumaccess}
\bibliographystyle{ieeetran}

\end{document}